\documentstyle[11pt]{article}

\begin{document}
\centerline{\LARGE\bf MASSIVE SUPERPARTICLE}
\medskip
\centerline{\LARGE\bf WITH TENSORIAL CENTRAL CHARGES}
\bigskip
\centerline{\Large {\bf S. Fedoruk}$^1$ and {\bf V. G. Zima}$^2$}
\smallskip
\begin{center}
$^1$ {\it Ukrainian Engineering--Pedagogical Academy,\\
     61003 Kharkiv, 16 Universitetska Str., Ukraine\\
     e-mail: fed@postmaster.co.uk} \\
$^2$ {\it Kharkiv National University,\\
     61077 Kharkiv, 4 Svobody Sq., Ukraine\\
     e-mail: zima@postmaster.co.uk}
\end{center}
\vspace{3cm}
\begin{abstract}
We construct the manifestly Lorenz-invariant
formulation of the $N=1$ $D=4$ massive superparticle with tensorial
central charges. The model contains a real parameter $k$ and
at $k\ne 0$ possesses one $\kappa$-symmetry while at $k=0$
the number of $\kappa$-symmetry is two.
The equivalence of the formulations at all $k\ne 0$ is obtained.
The local  transformations of $\kappa$-symmetry are written out.
It is considered the using of index  spinor
for construction of the tensorial central charges.
It is obtained the equivalence
at classical level between the massive $D=4$ superparticle with
one $\kappa$-symmetry and the massive $D=4$ spinning particle
\end{abstract}

\newpage

\section{Introduction}

Last time there is the great interest in the analysis of the supersymmetric
models possessing supersymmetry with additional nonscalar central
charges~\cite{AzG}-\cite{GaHa}. Although the tensorial central charges in
the supersymmetry algebra are associated with topological contributions
of the superbrane theories it is attractive to obtain the superparticle
models having this symmetry. Recently such supersymmetric particle models
were obtained in massless case~\cite{BandL}, for $D=4$ with two or three
local $\kappa$-symmetries.

In present work we construct the model of the massive $N=1$ $D=4$
superparticle with tensorial central charges possessing one or two local
$\kappa$-symmetries~\footnote{The Lagrangian of the massive superparticle
with vector central  charge and with two $\kappa$-symmetries has been
presented already in~\cite{ZimFedC}}. In such a way we obtain in usual
space-time dimensions $D=4$ the superparticle with a single
$\kappa$-symmetry which is equivalent physically to the usual spinning
(spin $1/2$) particle~\cite{Pseudo,Pseu} in the positive energy sector.

In the pseudoclassical approach the Lagrangian of spinning particle has
the following form~\cite{Pseudo,Pseu}
\begin{equation} \label{L1/2}
L_{1/2}=p^\mu \dot x_\mu+
\frac{i}{2}(\psi^\mu \dot\psi_\mu +\psi_5 \dot\psi_5 )
-\frac{e}{2}(p^2+m^2)-i\chi (p\psi +m\psi_5 ) \, .
\end{equation}
The spin variables in this description are the Grassmannian (pseudo)vector
$\psi_\mu$ and the Grassmannian (pseudo)scalar $\psi_5$. Besides mass
constraint $T\equiv p^2 +m^2 \approx 0$ in Hamiltonian formalism the physical
sector of the model is subjected to the Grassmannian constraints from which
one Dirac constraint
\begin{equation} \label{D}
D \equiv p^\mu \psi_\mu +m\psi_5 \approx 0
\end{equation}
plays the role of the first class constraint and five self-conjugacy
condition for the Grassmannian variables
\begin{equation} \label{g}
g^\mu \equiv p_\psi^\mu -\frac{i}{2} \psi^\mu \approx 0 \, , \qquad
g_5 \equiv p_{\psi 5} -\frac{i}{2} \psi_5 \approx 0
\end{equation}
are the second class constraints. Thus the number of physical Grassmannian
degrees of freedom in the model~(\ref{L1/2}) is [number of
($\psi_\mu$, $\psi_5$, $p_{\psi\mu}$, $p_{\psi 5}$)] --
[number of the second class constraints ($g_\mu$, $g_5$)] --
2[number of the first class constraint ($D$)] = $3$.

The usual model of the massive CBS superparticle~\cite{CBS} with Grassmannian
spinor coordinates $\theta^\alpha$, $\bar\theta^{\dot\alpha}$
has only the fermionic spinor constraints
$$
d_{\theta\alpha} \equiv -ip_{\theta\alpha}-(\hat p\bar\theta)_\alpha
\approx 0 \, , \qquad
\bar d_{\theta\dot\alpha} \equiv -i\bar p_{\theta\dot\alpha}
-(\theta\hat p)_{\dot\alpha} \approx 0
$$
which all are the second class constraints. Here the number of the physical
Grassmannian degrees of freedom is
[number ($\theta^\alpha$, $\bar\theta^{\dot\alpha}$,
$p_{\theta\alpha}$, $\bar p_{\theta\dot\alpha}$)] --
[number ($d_\theta$, $\bar d_\theta$)] = $4$.
In order to obtain desired three physical fermionic degrees of freedom
it is necessary that from fermionic four spinor constraints three constraints
are of the second class whereas one constraint should be of the first class.
Such situation with nonsymmetric separation of the fermionic constraints
into the ones of first and second class has been proposed in massless
superparticle models~\cite{BandL} as well as in the massive particle
case~\cite{DelIvKr}. Precisely the situation with one first class
fermionic constraint has been present in~\cite{DelIvKr} in the construction
of $N=4 \to N=1$ PBGS in $d=1$. The relation between that model and our one
will be given below. Thus in the massive case the equivalence of spinning
particle and superparticle with tensorial central charges with one
$\kappa$-symmetry is expected. Let us note that in massless
case~\cite{VolZ,STV} the spinning particle is equivalent, at least on
classical level, to the usual CBS superparticle without any central charges.
This fact of identifying the local fermionic invariances of spinning particle
and $\kappa$-symmetries of superparticle is essential for superfield
formulation of massless superparticle theory~\cite{VolZ,STV}
and consequent generalizations on superbranes~\cite{BSV}.

Accounting above mentioned preliminary arguments for the possible
relation between massive spinning particle and massive superparticle
with tensorial central charges we take the following way for construction
of the superparticle model. We shall realize the covariant transition,
under preservation of the physical contents, from the model of the massive
spinning particle to the system with Grassmannian spinor variables.
As result of this procedure we arrive at model of the $N=1$ $D=4$ massive
superparticle with tensorial central charges possessing one local fermionic
invariance ($\kappa$-symmetry).

Covariant transition from the Grassmannian vector $\psi_\mu$ and scalar
$\psi_5$ to the Grassmannian spinors $\theta^\alpha$,
$\bar\theta^{\dot\alpha}$ requires using of the commuting spinor variables.
We introduce dynamical commuting spinor variables $\zeta^\alpha$,
$\bar\zeta^{\dot\alpha}=\overline{(\zeta^\alpha)}$.
Their canonical conjugate momenta are  $v_\alpha$, $\bar v_{\dot\alpha}$.
Introduced spinors are subject to the condition
\begin{equation} \label{r}
r-j\equiv \zeta\hat p\bar\zeta -j \approx 0  \quad\qquad
(r\equiv \zeta\hat p\bar\zeta) \, .
\end{equation}
This constraint inherent in the index spinor
approach~\cite{ZimFedL,ZimFedC,ZimFedJ} gives us the completeness condition
$$
r\delta_\alpha^\beta =\zeta_\alpha (\bar\zeta\tilde p)^\beta +
(\hat p\bar\zeta)_\alpha \zeta^\beta
$$
for spinors $\zeta$, $\hat p\zeta$. Here matrix $\tilde p$ is the contraction
of the space-time momentum and $\sigma$-matrices  with upper spinor indices,
$\tilde p \equiv p^\mu \tilde\sigma_\mu $,
$\tilde p = (p^{\dot\alpha \alpha})$. Corresponding matrix with
lower indices is denoted by $\hat p$,
$\hat p = p^\mu \sigma_\mu= (p_{\alpha \dot\alpha})$.
Numerical constant $j\ne 0$
plays the role of ``classical spin'' in the index
spinor formalism~\cite{ZimFedL,ZimFedC,ZimFedJ}.
We assume that the dynamics of the bosonic spinor variables is determined by
the Lagrangian of the following form
\begin{equation} \label{Lbs}
L_{\rm b.s.}=\dot\zeta v +\bar v\dot{\bar\zeta}
-\lambda (\zeta\hat p\bar\zeta -j) \, .
\end{equation}
One can exclude the variable $\zeta$ using its equation of motion.
Thus we obtain the second order Lagrangian
$$
L^{(2)}_{\rm b.s.}=\Lambda^{-1} \left[ m^{-1} \dot v\hat p\dot{\bar v} +
+\Lambda^2 (j/m) \right]
$$
with $\Lambda\equiv m\lambda$. This Lagrangian describes the motion
of a point in complex two--dimensional space parametrized by the Weyl
spinor $v$. Canonically conjugate space parametrized by $\zeta$
is restricted by the constant~(\ref{r}) and is obviously isomorphic
to the compact group manifold $SU(2)$. Formally, the constant $j/m$
plays here the role of point ``mass''.

The total system which we consider as initial under transition
to Grassmannian spinors is in fact the sum of the two sectors coupled
through the space-time coordinates. One of these sectors is the usual
massive spinning particle with Lagrangian~(\ref{L1/2}) whereas
the second is the sector of the bosonic spinor with Lagrangian~(\ref{Lbs}).
Thus the Lagrangian of the initial system has the following form
\begin{eqnarray}
L & = & L_{1/2} + L_{\rm b.s.} \nonumber \\
  & = & p \dot x + \frac{i}{2}(\psi \dot\psi +\psi_5 \dot\psi_5 )
-\frac{e}{2}(p^2+m^2)-i\chi (p\psi +m\psi_5 ) \nonumber \\
  & & {} + \dot\zeta v +\bar v\dot{\bar\zeta}
-\lambda (\zeta\hat p\bar\zeta -j) \, .  \label{Laspin}
\end{eqnarray}
As result of the constraint $\zeta\hat p\bar\zeta = j$
the sign of the constant $j$ defines the sign of the energy.
In following we consider the positive energy sector where $j>0$.

In this paper we use the $D=4$ spinor conventions of~\cite{Wess}.

\section{Massive superparticle with tensorial central charges. Lagrangian}

The conversion of spinning particle model
described by the Grassmannian variables $\psi_\mu$, $\psi_5$
to the model with the Grassmannian spinor variables
$\theta^\alpha$, $\bar\theta^{\dot\alpha}$ is realized by the general
resolution~\cite{ZimFedC} of the form
\begin{equation} \label{ps}
\psi_\mu =r^{-1/2} (\theta\sigma_\mu \tilde p \zeta +
\bar\zeta\tilde p \sigma_\mu \bar\theta )
-m\rho \zeta\sigma_\mu \bar\zeta \, ,
\end{equation}
\begin{equation} \label{p5}
\psi_5 =r^{-1/2} m(\zeta\theta  + \bar\theta \bar\zeta )
+r\rho +\tilde\psi_5 \, .
\end{equation}
The initial Grassmannian variables $\psi_\mu$, $\psi_5$
(5 variables) are expressed in terms of two Grassmannian
scalars $\rho$, $\tilde\psi_5$ and three components of spinor $\theta$.
Just for projections of $\psi_\mu \equiv -\frac{1}{2}
\tilde\sigma_\mu{}^{\dot\alpha\alpha} \hat\psi_{\alpha\dot\alpha}$
in the basis formed by spinors
$\zeta^\alpha $, $(\bar\zeta\tilde p)^\alpha $
we have
\begin{equation} \label{pro}
\zeta\hat\psi\bar\zeta = 2r^{1/2}
(\zeta\theta +\bar\theta \bar\zeta ) \, , \quad
\bar\zeta\tilde p\hat\psi\tilde p\zeta = 2mr^2 \rho \, ,
\end{equation}
\begin{equation} \label{proj}
\zeta\hat\psi\tilde p\zeta = 2r^{1/2}
(\zeta\hat p \bar\theta ) \, , \quad
\bar\zeta\tilde p\hat\psi\bar\zeta = 2r^{1/2}
(\theta\hat p \bar\zeta ) \, ,
\end{equation}
where $\hat \psi = \psi^\mu \sigma_\mu$.
The fourth component of the spinor
\begin{equation} \label{phi}
\phi = i(\theta\zeta - \bar\zeta\bar\theta )
\end{equation}
does not  participate in the expression for $\psi$-variables.
The inversion of~(\ref{ps}), (\ref{p5}) and (\ref{phi}) looks as follows
$$
\theta_\alpha  =\frac{1}{4} r^{-3/2}
\left[ (\zeta\hat\psi\bar\zeta)(\hat p\bar\zeta)_{\alpha}
+2(\bar\zeta\tilde p\hat\psi\bar\zeta) \zeta_{\alpha }\right]
+\frac{i}{2} r^{-1}\phi (\hat p\bar\zeta)_{\alpha} \, ,
$$
$$
\bar\theta_{\dot\alpha}=\frac{1}{4} r^{-3/2}
\left[ (\zeta\hat\psi\bar\zeta)(\zeta\hat p)_{\dot\alpha}
+2(\zeta\tilde\psi\hat p\zeta) \bar\zeta_{\dot\alpha }\right]
-\frac{i}{2} r^{-1}\phi (\zeta\hat p)_{\dot\alpha} \, ,
$$
$$
\rho =\frac{1}{2m} r^{-2}
(\bar\zeta\tilde p\hat\psi\tilde p\zeta) \, , \quad
\tilde\psi_5 =\frac{1}{m}(p^\mu \psi_\mu +m\psi_5)
- (2mr)^{-1} (\zeta\hat\psi\bar\zeta) (p^2 +m^2) \, .
$$

In the new variables the Dirac constraint takes a simple form.
On mass shell $p^2 +m^2 =0$ we  have
\begin{equation} \label{Dn}
D=p\psi +m\psi_5 =m\tilde\psi_5 \approx 0 \, .
\end{equation}
Moreover, we can extract from the new variables a pure gauge degree of freedom
for fermionic local symmetry of the spinning particle~\cite{Pseudo,Pseu}
(world-line supersymmetry)
$$
\delta\chi =\dot\epsilon \, ,\quad
\delta e=-2i\epsilon\chi \, ,\quad
\delta\psi_\mu =-\epsilon p_\mu \, ,\quad
\delta\psi_5 =-\epsilon m \, ,\quad
\delta x_\mu =i\epsilon\psi_\mu \, .
$$
In the new variables this transformation takes the form
$$
\delta\theta_\alpha =
-\frac{1}{4}\epsilon r^{-1/2}(\hat p\bar\zeta)_\alpha \, ,\quad
\delta\bar\theta_{\dot\alpha} =
-\frac{1}{4}\epsilon r^{-1/2}(\zeta\hat p)_{\dot\alpha} \, ,
$$
$$
\delta\rho = -\frac{1}{2}\epsilon mr^{-1} \, ,\quad
\delta\tilde\psi_5 = -\frac{1}{2m}\epsilon (p^2 +m^2) \approx 0 \, .
$$
Thus, the only transformed are the variable $\rho$ and one component
of spinor $\theta$
$$
\delta (\theta\zeta +\bar\zeta\bar\theta) =\frac{1}{2}\epsilon r^{1/2} \, .
$$
Subsequently the combination
$\rho + mr^{-3/2}(\theta\zeta +\bar\zeta\bar\theta)$
of this component $\theta$ and $\rho$ is invariant under the
gauge transformations,
$\delta [\rho + mr^{-3/2}(\theta\zeta +\bar\zeta\bar\theta)]=0$,
whereas the variable
\begin{equation} \label{rh}
\rho - mr^{-3/2}(\theta\zeta +\bar\zeta\bar\theta)
\end{equation}
is the pure gauge degree of freedom,
$\delta [\rho - mr^{-3/2}(\theta\zeta +\bar\zeta\bar\theta)]=
-mr^{-1}\epsilon$.

Accounting the equation of motion for bosonic spinor $\dot\zeta =0$ and
substituting the resolving expressions~(\ref{ps}), (\ref{p5})
for $\psi_\mu$, $\psi_5$ in the Lagrangian~(\ref{Laspin})
we arrive at the Lagrangian
\begin{eqnarray}
L&=&p(\dot x -i\dot\theta\sigma\bar\theta
+i\dot\theta\sigma\dot{\bar\theta})
-im^2r^{-1}(\theta\zeta{}\bar\zeta\dot{\bar\theta} -
\dot\theta\zeta{}\bar\zeta\bar\theta )  \nonumber \\
&& {}+\frac{i}{2}r^2
\left[ \rho + mr^{-3/2}(\theta\zeta +\bar\zeta\bar\theta)\right]
\left[ \dot\rho +
mr^{-3/2}(\dot\theta\zeta +\bar\zeta\dot{\bar\theta})\right] \nonumber \\
&& {}+\frac{i}{2}r
\left[ \rho -mr^{-3/2}(\theta\zeta +\bar\zeta\bar\theta)\right]
\dot{\tilde\psi}_5 +\frac{i}{2}r \tilde\psi_5
\left[ \dot\rho -mr^{-3/2}(\dot\theta\zeta +
\bar\zeta\dot{\bar\theta})\right] \nonumber \\
&& {}+\frac{i}{2} \tilde\psi_5 \dot{\tilde\psi}_5
-im\chi \tilde\psi_5 -\frac{e}{2}(p^2 +m^2) \nonumber \\
&& {}+ \dot\zeta v +\bar v \dot{\bar\zeta} -
\lambda (\zeta\hat p\bar\zeta -j) \, .  \label{Lspin1}
\end{eqnarray}
It should be stressed that the equation $\dot\zeta =0$ for bosonic spinor,
which has been used for derivation of the Lagrangian~(\ref{Lspin1}),
is reproduced by the same Lagrangian~(\ref{Lspin1}).
As we see from the Lagrangian, the gauge variable~(\ref{rh})
$\rho - mr^{-3/2}(\theta\zeta +\bar\zeta\bar\theta)$
is the corresponding conjugate variable for $\tilde\psi_5$
which generates the local transformations.
The simpler gauge fixing condition for it
$$
\rho - mr^{-3/2}(\theta\zeta +\bar\zeta\bar\theta) =0
$$
gives us the possibility to resolve the scalar
$\rho$ in term  of spinor projection $(\theta\zeta +\bar\zeta\bar\theta)$.
We take the more general condition of  this type
\begin{equation} \label{gauge}
\rho - mr^{-3/2}(\theta\zeta +\bar\zeta\bar\theta) =
2(k-1)mr^{-3/2}(\theta\zeta +\bar\zeta\bar\theta)
\end{equation}
which is the gauge fixing condition at all $k$ except $k=0$.
At $k=0$ (\ref{gauge}) is reduced to the condition on gauge
invariant variable
$$
\rho + mr^{-3/2}(\theta\zeta +\bar\zeta\bar\theta) =0
$$
and of course it is not a gauge fixing.

Substituting in the Lagrangian~(\ref{Lspin1})
the constraint condition $\tilde\psi_5 =0$
(the equation of motion for the Lagrange  multiplier  $\chi$)
and the expression
\begin{equation} \label{rho}
\rho  = (2k-1)mr^{-3/2}(\theta\zeta +\bar\zeta\bar\theta)
\end{equation}
(following from the gauge fixing condition~(\ref{gauge}) )
we obtain the Lagrangian
\begin{eqnarray}
L & = & p \dot\omega_\theta +iZ_{\alpha\beta}\theta^\alpha \dot\theta^\beta +
i\bar Z_{\dot\alpha\dot\beta}\bar\theta^{\dot\alpha}
\dot{\bar\theta}^{\dot\beta}
+iZ_{\alpha\dot\beta} ( \theta^\alpha\dot{\bar\theta}^{\dot\beta}
- \dot\theta^\alpha \bar\theta^{\dot\beta} )
- \frac{e}{2}(p^2+m^2) \nonumber \\
&& {} +\dot\zeta v + \bar v \dot{\bar\zeta} -
\lambda (\zeta\hat p\bar\zeta -j) \, . \label{Lagr}
\end{eqnarray}
In this expression $\omega_\theta\equiv\dot\omega_\theta\,d\tau=
dx-id\theta\sigma\bar\theta + i\theta\sigma d\bar\theta$
is the usual $N=1$ superinvariant $\omega$-form.
The quantities $Z_{\alpha\beta} = Z_{\beta\alpha}$,
$\bar Z_{\dot\alpha\dot\beta} = \overline{(Z_{\alpha\beta})}$
and $Z_{\alpha\dot\beta} = \overline{(Z_{\beta\dot\alpha})}$
are expressed in terms of bosonic spinor $\zeta$
(for similar formula see~\cite{BandL})
\begin{equation} \label{Zchar}
Z_{\alpha\beta}=2k^2 m^2 j^{-1} \zeta_\alpha \zeta_\beta \, ,\qquad
Z_{\alpha\dot\beta}=(2k^2-1) m^2 j^{-1}
\zeta_\alpha \bar\zeta_{\dot\beta} \, .
\end{equation}
$Z_{\alpha\beta}$ and $\bar Z_{\dot\alpha\dot\beta}$ are tensor
central charges (types $(1,0)$ and $(0,1)$) and $Z_{\alpha\dot\beta}$ is
vector one (type $(1/2,1/2)$) for the $D=4$ $N=1$
supersymmetry algebra~\cite{AzG}-\cite{GaHa}.

The same result is obtained if we consider the connection of the
systems~(\ref{Laspin}) and (\ref{Lagr}) in the Hamiltonian formalism.
Precisely there is the canonical transformation which connect
the models with each other. Now in order to make equal the number of
Grassmannian variables in the models we introduce pure
gauge variable $\phi$ in the initial model of the spinning particle.
Its pure  gauge nature is achieved by the presence of
the first class constraint
\begin{equation} \label{pphi}
p_\phi \approx 0
\end{equation}
in the initial model. So  in  the canonical transformation
we imply that the term $p_\phi \dot\phi -\mu p_\phi$ is
added to the Lagrangian~(\ref{Laspin}). Here $\mu$ is Lagrange multiplier.
The resolution of $\phi$ in terms of the  spinors is given by
the expression~(\ref{phi}).

As the generating function of the canonical transformation from system
with coordinates $\psi_\mu$, $\psi_5$, $\phi$, $x^\mu$,
$\zeta^\alpha$, $\bar\zeta^{\dot\alpha}$ to the system with coordinates
$\theta^\alpha$, $\bar\theta^{\dot\alpha}$, $\rho$,
$\tilde\psi_5$, $x^{\prime\mu}$, $\zeta^{\prime\alpha}$,
$\bar\zeta^{\prime\dot\alpha}$ we take
\begin{eqnarray}
F & = & -p_\psi^{\mu}\psi_{\mu}(p_{\mu}, \zeta, \theta, \rho) -
p_{\psi 5}\psi_{5}(\zeta, \theta, \rho, \tilde\psi_{5}) -
p_{\phi }\phi(\zeta, \theta) \nonumber \\
&& {} + \zeta^\alpha v^\prime_{\alpha } +
\bar v^\prime_{\dot\alpha} \bar\zeta^{\dot\alpha} -
p^\mu x^\prime_\mu \, . \label{GFun}
\end{eqnarray}
Here the expressions for old variables in term of new ones from
the right hand side of the equations~(\ref{ps}), (\ref{p5}), (\ref{phi})
have been used.
That construction of the generating function~(\ref{GFun})
reproduces by definition of the canonical transformation
the resolution~(\ref{ps}), (\ref{p5}), (\ref{phi}) of the
initial Grassmannian coordinates in spinors
$\psi_{\mu}=-\partial_l F/\partial p_\psi^{\mu}$,
$\psi_{5}=-\partial_l F/\partial p_{\psi 5}$,
$\phi =-\partial_l F/\partial p_\phi $ and
leaves invariable bosonic spinor coordinates
$\zeta^{\prime\alpha} ={\partial F}/{\partial v^\prime_{\alpha }} =
\zeta^\alpha$,
$\bar\zeta^{\prime\dot\alpha}=
{\partial F}/{\partial \bar v^\prime_{\dot\alpha}}=\bar\zeta^{\dot\alpha}$
and the momentum vector
$p^\prime_\mu  =-{\partial F}/{\partial x^{\prime\mu}} = p_\mu$.
The expression of new Grassmannian  momenta in terms of initial ones are
$$
p_{\theta\alpha }= -\partial_r F/\partial\theta^{\alpha}
=r^{-1/2}(\sigma_\mu\tilde p\zeta)_\alpha p_\psi^\mu
- mr^{-1/2}\zeta_{\alpha} p_{\psi 5}
+i\zeta_{\alpha} p_{\phi} \, ,
$$
$$
\bar p_{\theta\dot\alpha }= -\partial_r F/\partial\bar\theta^{\dot\alpha}
=r^{-1/2}(\bar\zeta\tilde p\sigma_\mu)_{\dot\alpha} p_\psi^\mu
- mr^{-1/2}\bar\zeta_{\dot\alpha} p_{\psi 5}
-i\bar\zeta_{\dot\alpha} p_{\phi} \, ,
$$
$$
p_{\rho }= -\partial_r F/\partial\rho
=-m(\zeta\sigma_\mu\bar\zeta) p_\psi^\mu +r p_{\psi 5} \, , \quad
p_{\tilde\psi 5} =-\partial_r F/\partial\tilde\psi_5 =p_{\psi 5} \, .
$$
The expressions of the initial bosonic spinor momenta
$v_\alpha = \partial F/\partial\zeta^\alpha$,
$\bar v_{\dot\alpha} = \partial F/\partial\bar\zeta^{\dot\alpha}$
and space-time coordinate $x_\mu = -\partial F/\partial p^\mu$
in terms of the new phase space coordinates contain
besides corresponding new phase variables the additional terms
depending on the new Grassmannian phase space variables.
These terms arise because of the dependence of the resolution
expressions~(\ref{ps}), (\ref{p5}), (\ref{phi}) on
$\zeta$, $\bar\zeta$ and $p$. Here we do not need in the explicit form
the expressions  for $v^\prime$,$\bar v^\prime$ and $x^\prime$
due to independence of all constraints on these phase variables.

Now we eliminate the variables $\tilde\psi_5$, $p_{\tilde\psi 5}$
by means of the Dirac constraint~(\ref{D}) and
gauge fixing condition for Dirac constraint
\begin{equation} \label{gfix}
p_{\tilde\psi 5} -i(k-1)mr^{-1/2}
\left[ \theta\zeta +\bar\zeta\bar\theta  \right] \approx 0
\end{equation}
at $k\ne 0$ \footnote{The diagonalized Dirac constraint
$D^\prime \equiv D - ip^\mu g_\mu -im g_5 =
-i[ p_\mu (p_\psi^\mu +\frac{i}{2} \psi^\mu ) +
m (p_{\psi 5} +\frac{i}{2} \psi_5 )]\approx 0$ has in new variables the form
$D^\prime =\frac{i}{4}r^{-1/2}
\left[ \bar\zeta\tilde p p_\theta +\bar p_\theta\tilde p \zeta\right]
-\frac{i}{2}mr^{-1}p_\rho +\frac{1}{2}m{\tilde\psi_5} \approx 0$.
The Poisson bracket of the condition~(\ref{gfix}) and $D^\prime$
is equal to $(km)/2$, i.e. at $k=0$ the condition~(\ref{gfix})
does not fix the gauge for the  Dirac constraint.}.
After  fulfilment of the additional canonical transformation
$p_\rho \to p_{\rho^\prime}=p_\rho -ikmr^{1/2}
\left[ \theta\zeta +\bar\zeta\bar\theta \right]$, which leads to resolving
form $p_{\rho^\prime} \approx 0$ of one Fermi-constraint from~(\ref{g}),
we eliminate the
variables $\rho$, $p_\rho$ with the help of two from five second class
Fermi-constraints~(\ref{g}). Because of the resolving form
of the constraints with respect to eliminated variables,
$\tilde\psi_5 \approx 0$ and $p_{\rho^\prime} \approx 0$,
the Dirac brackets for remaining variables are the same as their
Poisson brackets. After that the remaining Grassmannian constraints
take  the following form
\begin{equation} \label{psv1}
\bar\zeta\tilde p p_\theta - \bar p_\theta\tilde p \zeta \approx 0 \, ,
\end{equation}
\begin{equation} \label{psv2}
\left[ \bar\zeta\tilde p p_\theta + \bar p_\theta\tilde p \zeta \right]
- 4ik^2 m^2 \left[ \theta\zeta +\bar\zeta\bar\theta\right] \approx 0 \, ,
\end{equation}
\begin{equation} \label{psv3}
\zeta \left[ -ip_\theta -\hat p \bar\theta \right] \approx 0 \, , \qquad
\left[ -i\bar p_\theta -\theta\hat p \right] \bar\zeta \approx 0
\end{equation}
which are the same as the projections on spinors $\zeta$, $\hat p\bar\zeta$
of the Grassmannian spinor constraints
\begin{equation} \label{thcon1}
d_{\theta\alpha} \equiv -ip_{\theta\alpha}- (\hat p\bar\theta)_\alpha -
\theta^\beta Z_{\beta\alpha}-
Z_{\alpha\dot\beta}\bar\theta^{\dot\beta} \approx 0 \, ,
\end{equation}
\begin{equation} \label{thcon2}
\bar d_{\theta\dot\alpha} \equiv -i\bar p_{\theta\dot\alpha}-
(\theta\hat p)_{\dot\alpha} -
\bar Z_{\dot\alpha\dot\beta} \bar\theta^{\dot\beta}-
\theta^{\beta} Z_{\beta\dot\alpha} \approx 0
\end{equation}
with quantities $Z_{\alpha\beta}$, $Z_{\alpha\dot\beta}$
defined in~(\ref{Zchar}).
From invariance of the variables $\zeta^\alpha$, $\bar\zeta^{\dot\alpha}$,
$p_\mu$ under the canonical transformation all bosonic constraints, i.e.
$p^2 +m^2 \approx 0$ and $\zeta\hat p\bar\zeta -j\approx 0$, are not changed.
The system with remaining variables and the constraints is described
by the above mentioned Lagrangian~(\ref{Lagr}). The Lagrangian~(\ref{Lagr})
reproduces accurately this set of the constraints and nothing else.

Thus we establish that the model described by Lagrangian
$L=L_{1/2} + L_{\rm b.s.}$ is equivalent physically to the model with
Lagrangian $L=L_{\rm super} + L_{\rm b.s.}$ at classical level.
Here $L_{1/2}$ is the Lagrangian~(\ref{L1/2}) of the massive spinning particle
(spin $1/2$) whereas $L_{\rm super}$ is Lagrangian
of the massive $N=1$ superparticle with tensorial
central charges~(\ref{Zchar})
\begin{equation} \label{Lagr5}
L_{\rm super} =  p \dot\omega_\theta +
iZ_{\alpha\beta}\theta^\alpha \dot\theta^\beta +
i\bar Z_{\dot\alpha\dot\beta}\bar\theta^{\dot\alpha}
\dot{\bar\theta}^{\dot\beta}
+iZ_{\alpha\dot\beta} ( \theta^\alpha\dot{\bar\theta}^{\dot\beta}
- \dot\theta^\alpha \bar\theta^{\dot\beta} )
- \frac{e}{2}(p^2+m^2) \, .
\end{equation}
Lagrangians $L_{\rm b.s.}$ of the bosonic spinor in the both equivalent
models are quite identical.

It should be noted that the value of constant $k$
in the formula~(\ref{Zchar}) for central charges
of the superparticle is nonzero, $k \ne 0$, in the case of the equivalence
to the spinning particle. But in general the value $k=0$ is not forbidded
in model of superparticle with central charges. Next we consider the cases
both with $k \ne 0$ and $k=0$. As we see below at $k \ne 0$ and $k=0$
we have superparticle models with one and two $\kappa$-symmetries
respectively.

Alternative way for a proof of classical equivalence  of
the massive spin $1/2$ particle~(\ref{L1/2})
and the massive superparticle
with central charges~(\ref{Lagr}), at $k \ne 0$,
possessing one $\kappa$-symmetry is the reduction of both model to
physical degrees of freedom~\cite{Towns}.
In the examining positive energy sector
after the choice of gauge $\psi_- =\psi_0 -\psi_5 =0$
for Dirac constraint and exclusion of $\psi_+ =\psi_0 +\psi_5$
by means of the constraint condition we obtain for the
physical Grassmannian degrees of freedom of spinning
particle~\cite{GaGoTo,Towns} the Lagrangian in the form of
$L^{({\rm ph})}_{1/2,{\rm Gr}}=\frac{i}{2}\vec\psi\dot{\vec\psi}$.
On the other hand the Grassmannian part of the superparticle
Lagrangian $L_{\rm super}$ takes the form
$$
L^{({\rm ph})}_{{\rm super},{\rm Gr}}= i\bar q\dot q -
iq\dot{\bar q} +2k^2 i\eta\dot\eta
$$
after using of the variables
\begin{equation} \label{nov1}
\eta =mr^{-1/2}(\theta\zeta +\bar\zeta\bar\theta) \, ,\quad
\sigma =-imr^{-1/2}(\theta\zeta -\bar\zeta\bar\theta) \, ,
\end{equation}
\begin{equation} \label{nov2}
q= r^{-1/2} (\theta\hat p\bar\zeta)\, ,\quad
\bar q= r^{-1/2} (\zeta\hat p\bar\theta) \, .
\end{equation}
Setting
$$
q=(\psi_1 +i\psi_2 )/2 \, ,\quad \bar q=(\psi_1 -i\psi_2 )/2 \, ,\quad
\eta = \psi_3 /2k
$$
we obtain exactly the same Grassmannian part of the Lagrangian
\begin{equation} \label{Lph}
L^{({\rm ph})}_{{\rm super},{\rm Gr}}=
L^{({\rm ph})}_{1/2,{\rm Gr}}=\frac{i}{2}\vec\psi\dot{\vec\psi}  \, .
\end{equation}

Such Lagrangian for the physical Grassmannian variables comes out also
from work~\cite{DelIvKr} in non-Lorentz covariant Grassmannian sector
$N=4 \to N=1$ PBGS. In first order formalism
the target space action of this work has the Lagrangian
\begin{equation} \label{Ivan}
L=\vec P \vec\Pi -P^0 \Pi^0 +\frac{e}{2}(P^{0{}2}-\vec P^2 -1)
-\Theta\dot\Theta -\vec\Psi\dot{\vec\Psi}
\end{equation}
where $\Pi^0 =\dot X^0 +\Theta\dot\Theta +\vec\Psi\dot{\vec\Psi}$,
$\vec\Pi =\dot{\vec Y} -\dot\Theta\vec\Psi +\Theta\dot{\vec\Psi}$
(we remain here the notations of~\cite{DelIvKr}). In accounting the last
expressions, the Lagrangian~(\ref{Ivan}) takes the form
$$
L=\vec P \dot{\vec Y} -P^0 \dot X^0 +\frac{e}{2}(P^{0{}2}-\vec P^2 -1)
-(P^0 +1)\left[ \vec\Psi - \frac{1}{P^0 +1}\vec P\Theta \right]
\left[ \vec\Psi - \frac{1}{P^0 +1}\vec P\Theta \right]^\cdot \, .
$$
After using of the variables
$$
\vec\psi = \sqrt{2}(P^0 +1)^{1/2}
\left[ \vec\Psi - \frac{1}{P^0 +1}\vec P\Theta \right]
$$
we obtain exactly the Lagrangian~(\ref{Lph}) for Grassmannian variables.

In order to analyse the properties of the obtained massive superparticle
with tensorial central charges let us consider the model  of spinning
particle with index spinor~\cite{ZimFedL,ZimFedC,ZimFedJ} as additional
bosonic coordinates. It is naturally because we have used
for bosonic  spinor the relation $\zeta\hat  p\bar\zeta -j \approx 0$
which is inherent in the index  spinor  approach.
In  the Hamiltonian formalism the index spinor sector is restricted
by the spinor self-conjugacy conditions
\begin{equation} \label{dzeta}
d_\zeta\equiv ip_\zeta-\hat p\bar\zeta\approx 0\, ,\qquad
\bar d_\zeta\equiv -i\hat p_\zeta-\zeta\hat p \approx 0
\end{equation}
which are the second  class constraints  in the massive case.
It is  achieved in above model~(\ref{Laspin}) by the substitution
$v=-i\hat p\bar\zeta$, $\bar v=i\zeta\hat p$.
Then $L_{\rm b.s.}$~(\ref{Lbs}) takes the form
of the index spinor Lagrangian~\cite{ZimFedL}
\begin{equation} \label{ind}
L_{\rm index} = -i\dot\zeta\hat p{\bar\zeta} +
i\zeta\hat p\dot{\bar\zeta} -\lambda (\zeta\hat p{\bar\zeta}-j)\, .
\end{equation}
Included in the Lagrangian the constraint $\zeta\hat  p\bar\zeta -j \approx 0$
generates in Hamiltonian formalism the spin constraint
\begin{equation} \label{consp}
\frac{i}{2}(\zeta p_\zeta -\bar p_\zeta\bar\zeta)-j \approx 0
\end{equation}
which together
with second class constraints~(\ref{dzeta}) lead~\cite{ZimFedL} to the
particle state of the single spin associated with given sector of index
spinor. Spin of the particle in the quantum spectrum is the
value of the constant $j$ renormalized by ordering constants
(thus $j$ can be named ``classical spin'').

The realization of the previously considered canonical transformation
to the model with Lagrangian $L^\prime =L_{1/2} + L_{\rm index}$,
i.e. $L_{\rm index}$ instead $L_{\rm b.s.}$ in~(\ref{Laspin}),
leads to the Lagrangian
\begin{eqnarray}
L^\prime & =  p \dot\omega  & +iZ_{\alpha\beta}\theta^\alpha \dot\theta^\beta +
i\bar Z_{\dot\alpha\dot\beta}\bar\theta^{\dot\alpha}
\dot{\bar\theta}^{\dot\beta}
+iZ_{\alpha\dot\beta} ( \theta^\alpha\dot{\bar\theta}^{\dot\beta}
- \dot\theta^\alpha \bar\theta^{\dot\beta} ) \nonumber \\
&&{} +iY_{\alpha\beta}\zeta^\alpha \dot\zeta^\beta +
i\bar Y_{\dot\alpha\dot\beta}\bar\zeta^{\dot\alpha}\dot{\bar\zeta}^{\dot\beta}
+iY_{\alpha\dot\beta} ( \zeta^\alpha\dot{\bar\zeta}^{\dot\beta}
+ \dot\zeta^\alpha \bar\zeta^{\dot\beta} ) \nonumber \\
&&{} -iN (\dot\zeta\hat p\bar\zeta -\zeta\hat p\dot{\bar\zeta}) \nonumber \\
&&{} - \frac{e}{2}(p^2+m^2) -\lambda (\zeta\hat p\bar\zeta -j) \, .
\label{Lprime}
\end{eqnarray}
Here the form $\omega\equiv\dot\omega\,d\tau=dx-id\zeta\sigma\bar\zeta+
i\zeta\sigma d\bar\zeta-id\theta\sigma\bar\theta +i\theta\sigma d\bar\theta$
is invariant with respect to the transformations of the usual $N=1$
supersymmetry with Grassmannian spinor parameter and ``bosonic supersymmetry''
with $c$-number spinor parameter~\cite{ZimFedL,ZimFedC,ZimFedJ}.
The central charges $Z_{\alpha\beta}$,
$Z_{\alpha\dot\beta}$ have the same form~(\ref{Zchar}).
So the kinetic terms of the space-time coordinate and Grassmannian
spinor in $L^\prime$ (\ref{Lprime}) are identical to the corresponding
terms in $L$ (\ref{Lagr}) and hence the algebras of the fermionic
constraints  in both models are identical. But the kinetic terms
of the index spinor in Lagrangian $L^\prime$ are different
from the kinetic terms of the bosonic spinor in Lagrangian $L$
by additional terms with quantities
$$
Y_{\alpha\beta} = 2k(k-2) m^2 j^{-1}\theta_\alpha\theta_\beta \, ,\qquad
\bar Y_{\dot\alpha\dot\beta} = -\overline{(Y_{\alpha\beta})}  \, ,
$$
\begin{equation}
Y_{\alpha\dot\beta}=
-(2k^2-4k+1) m^2 j^{-1} \theta_\alpha\bar\theta_{\dot\beta}
\end{equation}
which can be regarded as the central charges of the ``bosonic SUSY''
as well as
\begin{equation}
N\equiv j^{-1} \left[ (\theta\hat p \bar\theta )
+2(2k-1) m^2 j^{-1} (\theta\zeta) (\bar\zeta\bar\theta)\right] \, .
\end{equation}
The appearance of these extra terms is the result of modification of
index spinor momenta $p_\zeta$, $\bar p_\zeta$ under
the canonical transformation and, as consequence,
the modification of the spin constraint~(\ref{consp}) and
bosonic spinor constraints~(\ref{dzeta}) expressed by new variables.

Specific peculiarity of the model~(\ref{Lprime}) with index spinor
is an interconnection  between usual fermionic supersymmetry and
``bosonic one'' and at present its meaning is not yet quite clear.
Some duality appears in the invariance under permutation of Grassmannian
and bosonic spinors both $\omega$-form and certain terms with central charges
of different types.

\section{Massive superparticle with tensorial central charges. Invariances}

The massive superparticle~(\ref{Lagr}) with tensorial central charges
possesses the usual target space supersymmetry
\begin{equation}
\delta\theta^\alpha =\epsilon^\alpha \, , \quad
\delta\bar\theta^{\dot\alpha} =\bar\epsilon^{\dot\alpha} \, , \quad
\delta x_\mu =i\theta\sigma_\mu \delta\bar\theta
-i\delta\theta\sigma_\mu \bar\theta
\end{equation}
with constant Grassmannian parameter $\epsilon^\alpha$.
As usual in the cases  of the formulation without
central charge coordinates~\cite{AsLuk}  the Lagrangian $L$ is
quasi-invariant. With accounting of the bosonic spinor equation of motion
$\dot\zeta =0$ its  variation is the full derivative
\begin{equation}
\delta L=\left( iZ_{\alpha\beta}\epsilon^\alpha \theta^\beta
+iZ_{\alpha\dot\beta}\epsilon^\alpha{\bar\theta}^{\dot\beta} \right)^\cdot
{}+ \quad {\rm c.{}c.}
\end{equation}
Then the generators  of the supersymmetry transformations
$$
Q_\alpha =\frac{\partial}{\partial\theta^\alpha} +(\hat p\bar\theta)_\alpha
+ \theta^\beta Z_{\beta\alpha}+
Z_{\alpha\dot\beta}\bar\theta^{\dot\beta} \, ,
$$
\begin{equation}
\bar Q_{\dot\alpha} =\frac{\partial}{\partial\bar\theta^{\dot\alpha}} +
(\theta\hat p)_{\dot\alpha} +
\bar Z_{\dot\alpha\dot\beta} \bar\theta^{\dot\beta}+
\theta^{\beta} Z_{\beta\dot\alpha}
\end{equation}
contain  ``anomalous'' extra piece with central
charges~(\ref{Zchar})~\cite{AsLuk,AzG}.
The algebra of SUSY generators
\begin{equation}
\left\{ Q_\alpha , Q_\beta \right\} =2Z_{\alpha\beta} \, , \quad
\left\{ Q_\alpha , \bar Q_{\dot\beta} \right\}
=2(p_{\alpha\dot\beta} + Z_{\alpha\dot\beta})
\end{equation}
is the  $N=1$ $D=4$ SUSY algebra extended by tensorial central
charges~\cite{AzG}-\cite{GaHa}.

Of course we can  introduce the coordinates of central charges
introducing terms with derivatives of these coordinates
to the multipliers at central charges in~(\ref{Lagr})~\cite{AsLuk,BandL}.
Then the model becomes not only quasi-invariant but SUSY invariant.

The price for the presence of the supersymmetry is the infinite number
of the spin states in the spectrum. At the restriction of
the bosonic spinor sector to the index spinor one the number
of the states in spectrum becomes finite but the supersymmetry
disappears. But in both cases, (\ref{Lagr}) and (\ref{Lprime}),
the models possess local $\kappa$-symmetries.

For local transformation of the Grassmannian spinor
\begin{equation}
\delta \theta^\alpha =i\kappa (\bar\zeta\tilde p)^\alpha \, , \qquad
\delta \bar\theta^{\dot\alpha} =-i\bar\kappa (\tilde p\zeta)^{\dot\alpha}
\end{equation}
and standard Siegel transformation~\cite{AsLuk,Sieg} of  the
space-time coordinate
\begin{equation}
\delta x_\mu =-i\theta\sigma_\mu \delta\bar\theta
+i\delta\theta\sigma_\mu \bar\theta
\end{equation}
with local complex Grassmannian parameter $\kappa (\tau )$ the variation
of the Lagrangians up to a total derivative is
\begin{eqnarray}
\delta L & = &
-2k^2m^2(\theta\zeta +\bar\zeta\bar\theta)(\kappa -\bar\kappa)^\cdot
+2k^2m^2(\theta\zeta +\bar\zeta\bar\theta)^\cdot(\kappa -\bar\kappa)
 \nonumber \\
&&{} - 4km^2j^{-1}[(\theta\hat p\bar\zeta)\zeta\dot\zeta
+(\zeta\hat p\bar\theta)\dot{\bar\zeta}\bar\zeta](\kappa -\bar\kappa) \, .
\end{eqnarray}
As we see, $\delta L=0$ for real $\kappa =\bar\kappa$
at arbitrary values of constant $k$. But at $k=0$ we have $\delta L=0$
for arbitrary complex parameter $\kappa$.
Thus at $k\ne 0$ when the tensor central charge
$Z_{\alpha\beta}$ is present the models have one $\kappa$-symmetry with
real Grassmannian parameter $\kappa =\bar\kappa$.
But at $k=0$ when there is only the vector central charge
$Z_{\alpha\dot\beta}$ we have two $\kappa$-symmetries with complex
Grassmannian parameter $\kappa$.

A first class constraint is associated to each local invariance in
Hamiltonian formalism. As is already noted our systems are described by
the fermionic constraints (covariant derivatives)
(\ref{thcon1}), (\ref{thcon2}). Their Poisson brackets algebra is
$$
\left\{ d_{\theta\alpha}, d_{\theta\beta} \right\} =2iZ_{\alpha\beta}
\, , \quad
\left\{ \bar d_{\theta\dot\alpha}, \bar d_{\theta\dot\beta} \right\}
=2i\bar Z_{\dot\alpha\dot\beta} \, ,
$$
\begin{equation}
\left\{ d_{\theta\alpha}, \bar d_{\theta\dot\beta} \right\}
=2i\left( p_{\alpha\dot\beta}+ Z_{\alpha\dot\beta} \right)
\end{equation}
with central charges~(\ref{Zchar}).
Covariant separation of the fermionic first and second class
constraints is achieved by the projection
on the spinors $\zeta_\alpha$, $(\hat p\bar\zeta)_\alpha$. Let us put
\begin{equation} \label{conprc}
\chi_\theta \equiv \zeta d_\theta = -i\zeta p_\theta -
\zeta\hat p\bar\theta \approx 0 \, ,
\quad
\bar\chi_\theta \equiv \bar d_\theta \bar\zeta=
-i\bar p_\theta \bar\zeta -
\theta\hat p\bar\zeta  \approx 0 \, ,
\end{equation}
\begin{equation} \label{conprg}
g_\theta \equiv \bar\zeta\tilde p d_\theta +\bar d_\theta\tilde p\zeta =
-i(\bar\zeta\tilde p p_\theta + \bar p_\theta\tilde p\zeta )-
4k^2 m^2 (\theta\zeta +\bar\zeta\bar\theta) \approx 0 \, ,
\end{equation}
\begin{equation} \label{conprf}
f_\theta \equiv i(\bar\zeta\tilde p d_\theta -\bar d_\theta\tilde p\zeta )=
\bar\zeta\tilde p p_\theta - \bar p_\theta\tilde p\zeta \approx 0 \, .
\end{equation}
The nonzero Poisson brackets of these projections are
\begin{equation}
\left\{ \chi_{\theta}, \bar\chi_{\theta} \right\} =2ij \, , \qquad
\left\{ g_{\theta}, g_{\theta} \right\} =16 k^2 m^2 ij \, .
\end{equation}
Thus the constraints $\chi_{\theta}$, $\bar\chi_{\theta}$ are always
the second class constraints whereas the constraint $f_\theta$ is always
the first class constraint generating one $\kappa$-symmetry
with  local parameter $(\kappa +\bar\kappa )$ on variable
$(\theta\zeta -\bar\zeta\bar\theta)$,
$\left\{ f_{\theta}, \theta\zeta -\bar\zeta\bar\theta \right\} =2r$,
$\delta (\theta\zeta -\bar\zeta\bar\theta)=ir(\kappa +\bar\kappa )$.
The constraint $g_\theta$ is the second class constraint at $k\ne 0$.
But at $k=0$ the constraint $g_\theta$ becomes the first class constraint
and generates additional $\kappa$-symmetry with  local parameter
$i(\kappa -\bar\kappa )$ on variable $(\theta\zeta +\bar\zeta\bar\theta)$,
$\left\{ g_{\theta}, \theta\zeta +\bar\zeta\bar\theta \right\} =-2ir$,
$\delta (\theta\zeta +\bar\zeta\bar\theta)=ir(\kappa -\bar\kappa )$.

Thus we obtain the models of the $D=4$ $N=1$ massive superparticle with
tensorial central charges possessing one or two Siegel $\kappa$-symmetries.
In the language of the brane theories these models correspond to the BPS
superbrane configurations preserving $1/4$ or $1/2$ of supersymmetry
(see~\cite{GaHa} and references there).

It should be noted that constant $k$ in the construction of the superparticle
appears in the gauge fixing condition under transition from the
spinning particle. Therefore at all $k\ne 0$ the superparticle has quite
similar systems of the constraints and  the same  number of physical
degrees of freedom. The models at all $k\ne 0$ are equivalent.
Under transformations which can be considered as canonical transformations
\begin{equation}
\theta^\alpha \to \theta^\alpha +br^{-1}(\theta\zeta +\bar\zeta\bar\theta)
(\bar\zeta\tilde p)^\alpha \, , \quad
\bar\theta^{\dot\alpha} \to \bar\theta^{\dot\alpha} +
br^{-1}(\theta\zeta +\bar\zeta\bar\theta)(\tilde p\zeta)^{\dot\alpha}
\end{equation}
where $b$ is real number the Lagrangian $L$ (or $L^\prime$) transforms
into the same Lagrangian with $ak$ in place of  $k$ where $a\equiv 1+2b$.
As final result at level of the free superparticle we have two substantially
different models of the massive superparticle with tensorial central charges.
First of them at $k=1/\sqrt{2}$ has only tensor central charge  $Z_{\alpha\beta}$ and possesses
one $\kappa$-symmetry. Second model at $k=0$ has only vector central charge
$Z_{\alpha\dot\beta}$ and possesses two $\kappa$-symmetries.

\section{Conclusion}

In this work we presented the manifestly Lorenz-invariant
formulation of the $D=4$ $N=1$ free massive superparticle with tensorial
central charges. The model contains a real parameter $k$ and
at $k\ne 0$ it has one $\kappa$-symmetry while at $k=0$
the number of $\kappa$-symmetries is two.

In process of  the construction it is established the equivalence
at classical level between the massive $D=4$ $N=1$ superparticle with
one $\kappa$-symmetry and the massive $D=4$ $n=1$ spinning particle.
But they may lead to distinct quantum theories~\cite{Towns}.
Below we establish that the spinning particle and superparticle with
tensorial central  charges, which have  index spinor as additional one,
have identical  state spectrum.
By analogy with results in paper~\cite{Pseudo,Pseu,ZimFedL}
the first operator quantization of the spinning particle
with index spinor described by Lagrangian $L_{1/2} + L_{\rm index}$
is immediate. Wave function in the model  is defined by Dirac spinor with
(anti)holomorphic dependence in index spinor of homogeneity degree $2J$
where $J$ is the classical  spin $j$ renormalized by the ordering constant.
Writing  Dirac spinor in terms of Weyl spinors as
$\left( \psi \atop \chi \right)$, in according to analysis carried out
in~\cite{ZimFedL} we have in holomorphic case two multispinor fields
$\psi_{\alpha_1\ldots\alpha_{2J}\beta}$ and
$\chi_{\alpha_1\ldots\alpha_{2J}\dot\beta}$ which are symmetrical
in $2J$  indices  $\alpha$s. Here $\beta$ and $\dot\beta$ correspond
to bispinor index. These field  are connected with each other by
Dirac equation
\begin{equation}
\left( {0 \quad \tilde p} \atop {\hat p \quad 0} \right)
\left( \psi \atop \chi \right) = m \left( \psi \atop \chi \right)
\end{equation}
(quantum counterpart of the Dirac constraint~(\ref{D})).
Comparison with superparticle model is more immediate if we take the field
$\chi_{\alpha_1\ldots\alpha_{2J}\dot\beta}$ as basic one.
But the field $\psi_{\alpha_1\ldots\alpha_{2J-1}\alpha_{2J}\beta} =
\phi_{(\alpha_1\ldots\alpha_{2J}\beta)} +
\phi_{(\alpha_1\ldots\alpha_{2J-1}}\epsilon_{\alpha_{2J})\beta}$
exhibits simply that two spins $J\pm\frac{1}{2}$ are presented in spectrum
at fixed $J$ as it should be when one adds spin $J$ which is given by
index spinor and spin $\frac{1}{2}$ which corresponds to the Grassmannian
variables $\psi_\mu$, $\psi_5$ of the pseudoclassical mechanics under
quantization.

The quantization of the superparticle~(\ref{Lprime}) is suitable to carry out
in variables~(\ref{nov1}), (\ref{nov2}) in term of which
the fermionic constraints~(\ref{conprc})-(\ref{conprf}) takes
the extremely simple form
$$
ip_q +\bar q \approx 0 \, ,\quad
i\bar p_q + q \approx 0 \, ,
$$
\begin{equation}
ip_\eta +2k^2 \eta \approx 0 \, ,
\end{equation}
$$
p_\sigma \approx 0 \, .
$$
We gauging out the variable $\sigma$, the introduce the Dirac brackets
for taking into account of the  fermionic second class constraints and
the represent the remaining fermionic variables $q$, $\bar q$, $\eta$
(in fact $\vec\psi$) by means of the usual Pauli $\sigma$-matrices.
Thus the wave  function of this problem has two components
depending appropriately on index spinor and space-time variables.
The quantization of the bosonic spinor sector shows certain difference
with~\cite{ZimFedL}. Additional term of the form $q\bar q$  in spin
constraint~(\ref{consp}) arising due to interaction of bosonic and fermionic
sectors leads to different homogeneity degrees (which correspond to
different representations of Lorentz group)  for  two  components of wave
function. Bosonic spinor  constraints~(\ref{dzeta})  ((anti)homogeneity
conditions) acquire the additional  terms both with $q\bar q$ and also
$q\eta$ (or $\bar q\eta$). These  last  terms, which  are proportional
$\sigma_+$ (or $\sigma_-$), $\sigma_\pm \equiv (\sigma_1 \pm i\sigma_2)/2$
in matrix realization of Grassmannian variables, connect two components
of wave function. As result the  irreducible $(2J+1)$-component spinor
field $\phi_{\alpha_1\ldots\alpha_{2J+1}}$, in term of which one component
of wave function is determined, is expressed by Dirac equation
\begin{equation}
p_{\gamma\dot\beta}
\chi_{\alpha_1\ldots\alpha_{2J}}{}^{\dot\beta} =
m \phi_{\alpha_1\ldots\alpha_{2J}\gamma} \, .
\end{equation}
via field $\chi_{\alpha_1\ldots\alpha_{2J}\dot\beta}$ which determines
second component of wave function. This last field
$\chi_{\alpha_1\ldots\alpha_{2J}\dot\beta}$  can be identified  with
basic field of the spinning particle spectrum.

In case of models~(\ref{Laspin}) and (\ref{Lagr}), when there is not present
the truncation of bosonic spinor sector to the index one because of absence
of bosonic spinor constraints, the quantum equivalence apparently
remains too. One can expect it from the quite identity of bosonic sectors
of the models~(\ref{Laspin}) and (\ref{Lagr}) and identifying of physical
fermionic degrees of freedom which has been demonstrated in Sec. 2.

In case of the Lagrangian~(\ref{Lagr}) one can include vector central charge
$Z_\mu$ into vector  of space-time momentum by the shift
$p_\mu \to p_\mu + Z_\mu$ after taking into account
the bosonic spinor equation
of motion  $\dot\zeta =0$. Therefore  at $k=0$, when there  is vector central
charge only, it disappears completely from  the action  and superparticle
model reduces in fact to massless case.
Unlike this in the particle model~(\ref{Lprime}) with  index bosonic spinor
at $k=0$ the redefinition of momentum does not exclude vector central charge
due to accompanying modification of bosonic spinor and spin constraints.
In this case the wave function contains two usual spin-tensor fields
$\phi_{\alpha_1\ldots\alpha_{2J\pm 1}}$, satisfying massive Klein-Gordon
equation and disconnected with each other because of missing terms
with $q\eta$ in bosonic spinor constraints.

\medskip
We would like to thank I.A.Ban\-dos, E.A.Iva\-nov, S.O.Kri\-vo\-nos,
J.Lu\-kier\-ski, A.Yu.Nur\-ma\-gam\-be\-tov, D.P.So\-ro\-kin,
A.A.Zhel\-tu\-khin for interest to the work and for many useful discussions.
The authors are grateful to I.A.Ban\-dos, V.P.Be\-re\-zo\-voj,
A.Yu.Nur\-ma\-gam\-be\-tov, D.P.So\-ro\-kin
for the hospitality at the NSC Kharkov Institute of Physics and Technology.
This work was partially supported by research grant of the Ministry
of Education and Science of Ukraine.

\end{document}